# Model Engineering for Complex Systems

Jean Bézivin, Richard F. Paige, Uwe Aβmann, Bernhard Rumpe, Doug Schmidt

## 1. Introduction

Complex systems are hard to define [1]. Nevertheless they are more and more frequently encountered. Examples include a worldwide airline traffic management system, a global telecommunication or energy infrastructure or even the whole legacy portfolio accumulated for more than thirty years in a large insurance company. There are currently few engineering methods and tools to deal with them in practice. The purpose of this Dagstuhl Perspectives Workshop on Model Engineering for Complex Systems was to study the applicability of Model Driven Engineering (MDE) to the development and management of complex systems.

MDE is a software engineering field based on few simple and sound principles. Its power stems from the assumption of considering everything – engineering artefacts, manipulations of artefacts, etc – as a model [3].

Our intuition was that MDE may provide the right level of abstraction to move the study of complex systems from an informal goal to more concrete grounds.

In order to provide first evidence in support of this intuition, the workshop studied different visions and different approaches to the development and management of different kinds of complex systems. This note presents the summary of the discussions.

## 2. Complex systems

There are a number of examples of complex biological, ecological or societal complex systems discussed in the literature [5]. In the context of this note we are interested predominantly in Computer Based Complex Systems (CBCS), i.e. complex systems with a significant number of hardware or software components. These parts may be processing elements (processors, programs, processes, etc.) or data elements (memory, disks, repositories, files, etc.) or any kind of composite elements (hardware and software). One of the most important characteristics of a complex system is that it is composed of a very large number of individual parts. But there are also additional properties.

A CBCS is constantly in evolution with a past history, a present, and a future. This evolution is the consequence of the various interactions between the parts of the system. The evolution is permanent, i.e. the CBCS usually never stops, even when some parts are added, removed exchanged or under maintenance or repair.

A CBCS has a structure (or static architecture) and a dynamic behaviour. It is composed of elements that may themselves be CBCSs (with structure and behaviour) and no limit exists on this deep nesting.

In addition to structure and behaviour, a CBCS also has a goal defining its purpose in the context in which it is operating. As previously stated, this also applies to any component of this system. Important information is also the metadata associated to any component. The categories of metadata are quite diverse.

Another dimension of a CBCS is engineering heterogeneity. Many components are hardware and software elements produced in the last fifty years, with different types of technologies. For example many different hardware technologies, programming languages, APIs, operating systems, database organizations, network protocols, standards, or normative specifications have been used. Furthermore there may be a penalty to the use of any technology. This is often called *accidental* complexity by Fred Brooks, and it adds an artificial portion to the *essential* complexity of the base problem. Managing the accidental complexity accumulated by many layers of technological legacy is an important challenge in the management of CBCSs.

A CBCS is also often a distributed system, i.e. its elements are located on many widely dispersed physical locations.

By definition a CBCS may not be understood by one unique human operator. On the contrary many stakeholders will have different views on the system. These stakeholders may play different roles (architect, designer, implementer, maintainer, manager, user, etc.)

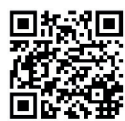

The interactions between the different parts of a CBCS are not random interactions and they follow specific patterns. Such a system is also characterized by the relationships that hold between its parts. Very often these relationships are informally characterized but in some occasions they may be explicitly represented. In either case they are quite important.

Managing a CBCS means observing, understanding, and controlling it. However management may imply additional operations like designing it, constructing it, measuring it, managing it, maintaining it, and many more. We are interested here in the support MDE may bring to all these operations on CBCS.

An additional property of CBCS is emergence. Emergence is the way complex systems and patterns arise out of a multiplicity of relatively simple interactions. Emergence is central to the theories of integrative levels and of complex systems. Emergence particularly makes engineering CBCS challenging because the behaviour of the overall system is difficult, if not impossible to predict from the behaviour of the individual components and connectors that make it up.

Overall, the discussion at the workshop led to the identification of the following critical properties of a CBCS:

- Size: such systems are often large and may be constructed from multiple different viewpoints. There was much discussion at the workshop on the challenges of integrating and reconciling different views (and hence viewpoints) in the CBCS engineering process. Importantly, this discussion was also raised at the *Challenges in MDE Workshop* at *MoDELS/UML 2008* in Toulouse in September.
- Heterogeneity: as discussed earlier, CBCS are often built from different technologies. A particular challenge that was noted at the workshop was the ability to be able to *replace* CBCS elements at run-time, i.e., as the CBCS was attempting to accomplish its goals. A novel flavour of this kind of run-time adaptation was the situation where the replacement is heterogeneous, i.e., where software needs to be replaced by hardware elements, and vice versa.
- Distribution: also, as mentioned earlier; the challenges of managing and deploying distributed CBCS, and developing them in the first place – with widely distributed teams – was discussed in many situations in the workshop. It was also noted that many of the challenges associated with distribution were not new and restricted to CBCS or MDE, but were inherent issues.
- Dynamicity: it was noted that CBCS were often dynamic, in the sense that they would adapt and evolve due to changes in their environment, new inputs, and feedback. Dynamicity combined with distribution poses particular challenges for development. There was some discussion at the workshop on the tension between dynamicity and criticality: it is inherently difficult to predict the behaviour of dynamic systems, yet predictability is essential in order to verify and validate a critical CBCS. There was some discussion on the use of contracts and lightweight static analysis for supporting critical CBCS development.
- Autonomy: CBCS exist that are constructed from autonomous individual components that are themselves capable of carrying out function and attempting to achieve goals. As discussed at the workshop, MDE may be well suited to building individual components that behave in a predictable way, but at the moment it is not clear how to integrate autonomous components using MDE in a disciplined and predictable way, so as to achieve system level goals. That being said, it was noted that this was not a specific difficulty with MDE: rather, it was one of the major challenges in building CBCS.

## 3. Model Driven Engineering

In the previous section, we have listed some important characteristics of complex systems, and some of the challenges in addressing them that were identified during the presentations and discussion sessions at the Dagstuhl Perspectives workshop. In this section we summarise our MDE-specific discussions related to engineering of complex systems.

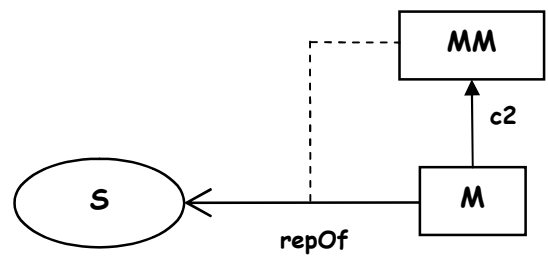

*Figure 1*. **The two basic relationships of MDE**

MDE considers models as first class citizens. A model is a representation of a system (relation *repOf* in Figure 1) and the nature of the model is defined by its metamodel. We say that a model conforms to its metamodel (relation *c2* in *Figure 1*).

MDE is mainly built on top of these two basic relations of *representation* and *conformance*, like object technology was mainly based on the relations of *instantiation* and *class inheritance*. MDE may be implemented with the help of object technology (or any other like functional). However the basic paradigms of MDE are inherently different from those of object technology.

Any system can be represented by a model. Then, we are able to give a homogeneous representation of a heterogeneous situation or phenomenon.

Metamodels are used as filters to define matter of interest in the system. Used as a typing system, they provide precise semantics to artefacts and relations between these artefacts. This homogeneity of definition of metamodels and models give us the power to apply operations on them in a generic way. Model to model transformations encode those operations.

A system can be "filtered" by more than one metamodel. As a complex system can not be understood and managed from one single point of view, being able to have different representations of this system is of great interest. For instance, we can imagine having a model of the static structure of a software application and a model of its execution trace (method calls, etc.) Moreover, as those representations are of the same system, they bear some relations (weaving models).

Model Driven Engineering provides some principles and tools to manage complex systems. But it is not sufficient by itself. The distribution and the handling of numerous artefacts, representation of complex systems as composition of artefacts that may be complex themselves are not addressed directly by MDE. These issues were discussed at the workshop.

### 3.1 Applying MDE to CBCS

The challenges of building CBCS that were identified during the workshop were summarized in Section 2, including dealing with aspects of size, heterogeneity, distributed, autonomy and dynamicity [16].

The connections between MDE and CBCS were discussed in detail in the latter part of the workshop. Many participants noted that some of the challenges noted were inherent to system engineering in general (particularly issues of size and distribution, and to a lesser extent heterogeneity). A number of participants pointed out specific solutions to challenges of heterogeneity and dynamicity from other communities.

For example, it was noted that the database community had well-defined principles and practices for handling heterogeneity in both databases and database management technology and that – at least conceptually – some of these ideas could be usefully applied to CBCS and integrated into state-of-the-art MDE technology as well. The debate on this issue seemed to focus on whether the MDE community understood how to efficiently model behaviour and semantics in a way that allowed their tooling to continue to be used.

A number of participants noted that one rich area for consideration in MDE for CBCS was in dealing with autonomous systems [11], [9]. Such systems – which may be self configuring, self healing, self optimizing – could apparently be at least partly addressed by current MDE techniques, but new ideas from the run-time systems management community (particularly for specifying "safe" or "acceptable" reconfigurations) were needed. Several participants from the more traditional complex systems community also noted that some of these kinds of systems were inherently challenging, if not impossible, to manage – e.g., bio-inspired systems – and that hoping to capture all the parameters of a "safe" or "acceptable" reconfiguration would be exceptionally difficult. One proposal for dealing with this was to exploit simulation technology – something that MDE can be used to support – to help to predict different possible reconfigurations that might arise, and to use simulation data to help manage, control, or at least direct the path of reconfiguration. This appears to be a challenging, yet likely fruitful direction of future research.

There was also discussion on industrial needs for MDE and CBCS. Participants discussed the benefits from having process support for MDE of CBCS (i.e., to support documenting and describing the engineering lifecycle), as well as challenges in integrating with legacy components and sub-systems (as discussed earlier). Technology for model understanding (or model "grokking") was presented that may help with this. Finally, the challenges of verification and validation of CBCS on an industrial scale were summarized and led to much debate that linked in to earlier discussions at the workshop on viewpoints and view integration [15,17]. The feeling was that individual verification technology (such as model checking, static analysis, or theorem proving) was insufficient and that integrated tool chains and workbenches for MDE were needed. Several examples of applying MDE to CBCS were discussed like megamodeling [2], [6], global traceability [8], model weaving [12], model merging [13], action semantics [14], etc.

## 4. Conclusion

The workshop was a success, and led to increased understanding of the fundamental characteristics of complex systems, the challenges of engineering them, the features for doing this that are currently offered by MDE, and several interesting future directions for research in MDE. Many participants noted that some of the challenges that were discussed had nothing to do with CBCS and MDE, but were simply challenges of building modern large-scale systems. However, it was acknowledged that MDE solutions for complex systems needed to address these challenges as well.

While the workshop was generally considered a success, many participants also noted that 2.5 days was simply too little time to fully discuss and start to collaborate on some of the key challenges. It is anticipated that some of the participants will work towards developing a successor Dagstuhl workshop, ideally of longer duration, for the future.

## 8. Acknowledgements

## 9. References

[1] Mikaël Barbero, Frédéric Jouault, Jean Bézivin: Model Driven Management of Complex Systems: Implementing the Macroscope's Vision. Pp.277-286, 15th Annual IEEE International Conference and Workshop on Engineering of Computer Based Systems (ECBS 2008), 31 March - 4 April 2008, Belfast, Northern Ireland. IEEE Computer Society 2008.

[2] Jean Bézivin, Frédéric Jouault, F, Patrick Valduriez: On the Need for Megamodels. In: Proceedings of the OOPSLA/GPCE: Best Practices for Model-Driven Software Development workshop, 19th Annual ACM Conference on Object-Oriented Programming, Systems, Languages, and Applications. 2004.

[3] Jean Bézivin: On the Unification Power of Models. *Software and System Modeling (SoSym)* 4(2):171—188. 2005.

[4] Frederick P. Brooks: *No Silver Bullet: Essence and Accidents of Software Engineering*. 1987.

[5] Joel De Rosnay: *The macroscope,* Harper & Row, New York, 1979.

[6] Rick Salay et al: An Eclipse-Based Tool Framework for Software Model Management, Eclipse Technology Exchange Workshop at OOPSLA 2007, Montreal, October 2008.

[7] Dimitrios S. Kolovos, Richard F. Paige, Fiona Polack: Detecting and Repairing Inconsistencies across Heterogeneous Models. ICST 2008: 356-364.

[8] Alek Radjenovic, Richard F. Paige: The Role of Dependency Links in Ensuring Architectural View Consistency. WICSA 2008: 199-208.

[9] Régine Laleau, Fiona Polack: Using formal metamodels to check consistency of functional views in information systems specification. *Information & Software Technology 50*(7-8): 797-814 (2008)

[10] Richard F. Paige, Phillip J. Brooke, Jonathan S. Ostroff: Metamodel-based model conformance and multiview consistency checking. *ACM Trans. Softw. Eng. Methodol. 16*(3): (2007).

[11] Andrew Weeks, Susan Stepney, Fiona Polack: Neutral Emergence and Coarse Graining. ECAL 2007: 1131-1140.

[12] Jean Bézivin, Salim Bouzitouna, Marcos Didonet Del Fabro, Marie-Pierre Gervais, Frédéric Jouault, Dimitrios S. Kolovos, Ivan Kurtev, Richard F. Paige: A Canonical Scheme for Model Composition. ECMDA-FA 2006: 346-360.

[13] Dimitrios S. Kolovos, Richard F. Paige, Fiona Polack: Merging Models with the Epsilon Merging Language (EML). MoDELS 2006: 215-229

[14] Richard F. Paige, Dimitrios S. Kolovos, Fiona Polack: An action semantics for MOF 2.0. SAC 2006: 1304-1305.

[15] C. Herrmann, H. Krahn, B. Rumpe, M. Schindler, S. Völkel: An Algebraic View on the Semantics of model Composition. ECMDA-FA 2007, LNCS 4530: 99-113

[16] R. France, B. Rumpe: Model-Driven Development of Complex Software: A Research Roadmap. Future of Software Engineering 2007 at ICSE: 37-54, IEEE, May 2007.

[17] D. Harel, B. Rumpe: Meaningful Modeling: What's the Semantics of "Semantics"? *IEEE Computer*: 37(10):64-72, October 2004.